%% file: hv97.tex
\documentstyle[aas2pp4,epsfig]{article}
 
\font\smcap=cmcsc10 

\def\hi{\smcap H$\,$i}

\def\simgt{$\gtrsim$}
\def\U{$U$}
\def\B{$B$}
\def\V{$V$}

\def\z{$z$}
\def\about{$\sim$}
\def\Mo{$M_{\odot}$}

\def\as{$^{\prime\prime}$}

\def\cf{{\it cf.}}

\slugcomment{Submitted to AJ}

\lefthead{Hibbard \& Vacca}
\righthead{On the Morphology of Intermediate Redshift Peculiars}

\begin{document}

\hfill{}
\bigskip
\bigskip
\bigskip
\bigskip
\title{The Apparent Morphology of Peculiar Galaxies \\
at Intermediate to High Redshifts}

\author{J.E.~Hibbard\altaffilmark{1,2} and W.D.~Vacca\altaffilmark{3}}
\affil{Institute for Astronomy , University of Hawaii,
    Honolulu, HI, 96822}

\altaffiltext{1}{Hubble Fellow}
\altaffiltext{2}{Present Address: NRAO, 520 Edgemont Rd., Charlottesville VA 22903}
\altaffiltext{3}{Beatrice Watson Parent Fellow}

\vfill
\centerline{To appear in {\it The Astronomical Journal}}

\bigskip\bigskip
\centerline{Electronic version of this paper is available at:}
\centerline {\it http://www.ifa.hawaii.edu/$\sim$hibbard/highZ/}
\centerline{or:}
\centerline {\it http://www.ifa.hawaii.edu/$\sim$vacca/highz.html}

\vfill
\noindent{\bf U{\sc NIVERSITY OF} H{\sc AWAII}}\hfill\break
{\bf INSTITUTE FOR ASTRONOMY}\hfill\break
{\bf 2680 Woodlawn Drive}\hfill\break
{\bf Honolulu, Hawaii 96822 USA}

\eject

\title{The Apparent Morphology of Peculiar Galaxies \\
at Intermediate to High Redshifts}

\author{J.E.~Hibbard\altaffilmark{1,2} and W.D.~Vacca\altaffilmark{3}}
\affil{Institute for Astronomy , University of Hawaii,
    Honolulu, HI, 96822}

\altaffiltext{1}{Hubble Fellow}
\altaffiltext{2}{Present Address: NRAO, 520 Edgemont Rd., Charlottesville VA 22903}
\altaffiltext{3}{Beatrice Watson Parrent Fellow}

\bigskip
\centerline{To be published in {\it The Astronomical Journal}}

\begin{abstract} 

To understand the range of morphologies observed in the galaxy
population at high redshift, it is instructive to evaluate the
appearance of local peculiar galaxies when placed at cosmological
distances.  To this end, we use rest frame ultraviolet (UV), $B$, and
$V$ band images of five nearby ($z<0.02$) interacting and/or
starbursting galaxies to simulate deep {\it HST} observations of
peculiar galaxies at medium to high redshifts.  In particular, we
simulate {\it Hubble Deep Field} (HDF) observations in the F606W and
F814W filters of starburst galaxies in the redshift range \z\about
0.5---2.5 by explicitly account for the combined effects of
band-shifting and surface brightness dimming. 

We find that extended morphological features remain readily visible in
the long exposures typical of the HDF out to redshifts of \about 1.  For
systems above $z\approx$1.5, the simulated morphologies look remarkably
similar to those of the faint objects found in the HDF and other deep
HST fields.  Peculiar starburst galaxies therefore appear to be the best
local analogs to the highest redshift galaxies in terms of morphology,
star formation rates, and spectral energy distributions.  Nevertheless,
photometric measurements of the $z>1.5$ images fail to recover the true
global properties of the underlying systems.  This is because the
high-$z$ observations are sensitive to the rest-frame UV emission, which
is dominated by the most active star forming regions.  The extended
distribution of starlight from more evolved populations would not
be detected.  Therefore, while the observed morphologies provide
information on the distribution of star formation within the systems,
they may reveal very little about their true dynamical states.  We conclude
that imaging observations in the restframe UV alone cannot reveal
whether high-\z\ systems (\z\simgt1.5) are proto-galaxies, proto-bulges,
or starbursts within a pre-existing population.  Definitive statements
regarding the global properties and dynamical states of these objects
require deep imaging observations at longer wavelengths.

\end{abstract}

\keywords{galaxies: morphology -- galaxies: evolution -- 
galaxies: starbursts -- galaxies: mergers -- galaxies: formation
-- cosmology: observations}

\twocolumn

\section{Introduction}

One of the most exciting advances in observational cosmology, made
possible by the superior spatial resolution of the Hubble Space
Telescope (HST), has been the identification of the objects comprising
the ``Faint Blue Excess" in the galaxy luminosity function with a
rapidly evolving population of galaxies with irregular and/or peculiar
morphologies at intermediate to high redshifts (e.g.~Colless et al.\
1994; Griffiths et al.\ 1994; Glazebrook et al.\ 1995; Driver et al.\
1995a,b; Cowie et al.\ 1995; Schade et al.\ 1996; Abraham et al.\ 1996;
van den Bergh et al.\ 1996; Odewahn et al.\ 1996).  Despite the numerous
recent studies of this high-redshift population, the true nature of
these objects remains elusive.  They may be higher redshift
manifestations of what are known locally to be disturbed and interacting
galaxies, or simply normal Hubble types which look peculiar due to the
effects of band-shifting and the loss of surface brightness sensitivity
(e.g.~Bohlin et al.\ 1991; Griffiths et al.\ 1994; Schade et al.\ 1995;
Giavalisco et al.\ 1996a; Williams et al.\ 1996; O'Connell \& Marcum
1996; Colley et al.\ 1996).  Alternatively, they may represent entirely
new classes of galaxies (e.g.~``Chain Galaxies", Cowie, Hu \& Songaila
1995; Koo et al.\ 1996; ``Tadpole Galaxies", van den Bergh et al.\ 1996)
or sub-galactic-sized objects in the process of forming galaxies
(``proto-galactic clumps'', Pascarelle et al.\ 1996; Steidel et al.\
1996; Lowenthal et al.\ 1997; Trager et al.\ 1997). 

To aid in the interpretation of the wealth of morphological data
provided by HST observations of high redshift galaxies, it is
instructive to first understand what nearby galaxies ($z \sim 0$) would
look like if they were placed at cosmological distances ($z > 1$).  Such
redshifting experiments have been conducted by several authors using a
range of normal Hubble types (Bohlin et al.\ 1991; Giavalisco et
al.\ 1996a; O'Connell \& Marcum 1996; Maoz 1997).  However, such samples
contain very few peculiar galaxies and even fewer systems with star
formation rates (SFRs) as high as those inferred for objects found at $z
> 2$ (Steidel et al.\ 1996; Giavalisco, Steidel \& Macchetto 1996b; Madau
et al.\ 1996; Cowie et al.\ 1997; Lowenthal et al.\ 1997; Trager et
al.\ 1997).  The only systems in the local universe with similarly high
star formation rates are luminous infrared (IR) and/or starburst
galaxies, the majority of which also exhibit very disturbed morphologies
(see e.g.~Sanders \& Mirabel 1996).  As such, these objects may be the
best analogs to the high-$z$ galaxies in terms of morphology and star
formation rates.  It therefore seems worthwhile to repeat the
redshifting experiments with a sample of interacting and/or starburst
galaxies. 

In this paper we use ultraviolet ($UV$), $B$ and \V-band images of five
nearby interacting and/or starburst galaxies to simulate their
appearance when placed at distances corresponding to redshifts between
0.5---2.5 and observed with HST.  In \S2 we describe the five systems
used in this study.  In \S3 we describe the data processing.  In \S4 we
discuss the technique for making the simulated observations.  In \S5 we
discuss the implications of these results for interpreting the
morphologies of high-redshift systems, and in \S6 we summarize our major
conclusions.  Throughout this work we adopt H$_o$=50 km s$^{-1}$
Mpc$^{-1}$ and q$_o$=0.5. 

\section{Peculiars at $z\approx 0$}

The peculiar galaxies used in this exercise were chosen on the basis of
the availability of data at the appropriate wavelengths.  In particular,
we chose systems with $B,V$ data available from an on-going imaging
survey of peculiar galaxies (Hibbard et al.\ in preparation) and with HST
FOC ($\lambda\approx 2320$ \AA) data available from the HST archives. 
All systems are optical starbursts and are known to contain large
numbers of Wolf-Rayet stars (Wolf-Rayet galaxies; Vacca \& Conti 1992). 
The FOC images were acquired as part of a larger study of the UV
properties of starburst and Wolf-Rayet galaxies.  The presence of large
numbers of Wolf-Rayet stars in these systems indicates that a
short-lived burst of massive star formation occurred within the last 7
Myr in each of these objects (Vacca \& Conti 1992; Schaerer \& Vacca
1997). 

The global properties of each system are given as the lowest redshift
entry in Table 1, and \B-band images (described below) are shown in
Figure 1.  Superposed on these images are the fields-of-view of the FOC
observations and labels of some features of interest.  While each of
these systems contains multiple optical condensations which might be
associated with separate progenitor nuclei, only in Arp 299 do the
components seem well enough separated to warrant individual catalog
designations (NGC 3690 and IC 694), and these are listed as separate
entries in Table 1.  All physical scales (distances, radii, and
luminosities) have been derived directly from heliocentric velocities
and the adopted Hubble constant of $H_o$=50 km s$^{-1}$ Mpc$^{-1}$. 

The first two systems, Arp 299 and NGC 1614, are optically luminous
($L_B \approx 1.9$ $L_B^*$ for Arp 299, 1.7 $L_B^*$ for NGC 1614;
Mazzarella \& Boroson 1993) \footnote{We adopt $M_B^*\approx$-21 as the
characteristic luminosity for a Schechter luminosity function; Loveday
et al.\ 1992.} and ultraluminous in the infrared ($\log
[L_{IR}/L_{\odot}]>12$).  Such extreme IR luminosities are believed to
originate primarily from UV photons that are quickly absorbed by dust
and reradiated in the infrared (e.g.  Sanders \& Mirabel 1996).  Arp 299
and NGC 1614 are two of the most active star forming systems in the
local universe, with inferred star formation rates ranging from 50 to
400 \Mo\ yr$^{-1}$ (Heckman, Armus \& Miley 1990; Storchi-Bergmann,
Calzetti \& Kinney 1994; Lan\c con \& Rocca-Volmerange 1996).  Both
systems appear to be on-going mergers between two disk galaxies of
normal Hubble types (Gehrz, Sramek \& Weedman 1983; Neff et al.\ 1990;
Hibbard \& Yun 1997), and both appear to be powered predominantly by
compact starbursts rather than AGN (Wynn-Williams et al.\ 1991; Condon
et al.\ 1991; Ridgeway, Wynn-Williams \& Becklin 1994; Vacca \& Conti
1992; Vacca et al.\ 1997).  NGC 1614 is a merger that is near
completion, with the two nuclei (labeled as {\bf A} and {\bf B} in
Fig.~1a) separated by 11\as\ = 5 kpc (Neff et al.\ 1990, Mazzarella \&
Boroson 1993).  Arp 299 is a less advanced merger composed of two
separate disk systems, IC 694 and NGC 3690, that are in contact but not
fully merged (see Fig.~1b).  Their nuclei (labeled {\bf A} and {\bf B}
in Fig.~1b) are separated by 20\as\ (7 kpc). 

The next two systems, NGC 1741 and He 2-10, are less active starburst
systems.  They are an order of magnitude less luminous in the IR, with
most of the energy from the starburst emerging in the optical and UV,
and they are slightly less luminous in the optical (0.8 $L_B^*$ for NGC
1741 and 0.6 $L_B^*$ for He 2-10).  The SFRs are high ($\sim 30 M_\odot
{\rm yr}^{-1}$), but not as large as in the other two systems.  NGC 1741
is a member of the Hickson Compact Group 31 (Hickson 1994).  Its unusual
morphology is clearly the result of an interaction or merger with other
members of the group.  For the purposes of this paper we include only
HCG 31a and HCG 31c (labeled {\bf A} and {\bf C} in Fig.~1c) in the
global measurements of NGC 1741.  He 2-10 (Fig.~1d) is a relatively
nearby Blue Compact Dwarf.  The reason for the prodigious massive star
formation activity in He 2-10 is not as clear as in the other three
cases, but {\hi} imaging suggests that a merger or accretion event is
occurring (Kobulnicky et al.\ 1995). 

The total star formation rates tabulated in Table 1 were computed by
comparing the luminosities in the various bands with those derived from
the Bruzual \& Charlot (1997) models, with the assumptions of a Salpeter
Initial Mass Function spanning the mass range 0.1--125 M$_\odot$ and
constant star formation with a duration of 10 Myrs.  The conversion
factors are given in the notes to the table.  At $z\approx0$, the total
star formation rates for the optical starbursts NGC 1741 and He 2-10
were derived from $L_B$; for NGC 1614 and Arp 299 the total SFR was
computed using the IR luminosity to represent the bolometric luminosity. 
In ultraluminous IR systems such as NGC 1614 and Arp 299 the IR
luminosity dominates the total energy output (Sanders \& Mirabel 1996),
and the use of $L_{IR}$ for $L_{bol}$ is a reasonable approximation. 

\section{Data Processing}

The \B- and \V-band images for Arp 299, NGC 1741, and NGC 1614 were
obtained with the University of Hawaii 88\as\ telescope and Tek 2048 CCD
at Mauna Kea Observatory in November 1996 and January 1997 under
photometric conditions and 1\as\ seeing.  The f/10 reimaging optics were
used, giving a plate scale of 0.22\as\ pixel$^{-1}$.  The total exposure
times were 1200 sec in \B-band and 900 sec in \V-band.  The data were
calibrated via observations of Landolt UBVRI standards (Landolt 1983)
observed on the same nights, with zeropoint errors of order 0.05 mag. 
The \B- and \V-band images of He 2-10 were obtained with the 0.9m
telescope at CTIO and are described in Corbin, Korista \& Vacca (1993). 
The data were corrected for foreground Galactic extinction using the
reddening law of Seaton (1979). 

The UV data are archival HST FOC images.  These images were acquired
before the HST servicing mission and therefore suffer from spherical
aberration.  The images of IC694 and NGC 1614 were processed with the
HST pipeline calibration routines.  The images of NGC 3690, NGC 1741,
and He 2-10 were corrected for pixel ``roll-over'' and then re-processed
with the HST calibration pipeline procedures.  Additional processing on
all images was performed to interpolate over blemishes and reseaux and
correct for flat-field non-linearity.  The exposure times for the FOC
images of He 2-10, NGC 1741, and NGC 1614 were each 996.75 s.  The final
images of NGC 3690 and IC 694 were constructed by averaging two
observations, each with exposure times of 895.75 s.  The HST FOC image
of NGC 1741 has been presented and discussed by Conti, Leitherer \&
Vacca (1996), who also analyzed a GHRS UV spectrum of one of the bright
starburst knots seen in this image.  A description of the FOC image of
He 2-10 can be found in Conti \& Vacca (1994); a description of the
images of NGC 3690 and IC 694 can be found in Vacca et al.\ (1997; see
also Meurer et al.\ 1995). 

Since the redshifting technique described below accounts for the changes
in the number of photons per wavelength interval per unit collecting
area as a function of redshift, there is some concern that the method,
although correct for extended sources, may be inappropriate for treating
the cosmological fading of unresolved sources (Colley et al.\ 1996). 
This will not be a problem as long as the width of the point spread
function (PSF) in the original images is mapped onto a region smaller
than the width of the PSF of the redshifted image.  For the ground-based
images the seeing is about the same as the full width at half maximum of
unresolved sources in the final ``drizzled'' images of the HDF
($\sim$3-4 pixels).  Therefore, as long as the rebinning factors (the
number of pixels in the original image that are mapped into a single
pixel of the final image) are larger than unity the smearing of
unresolved sources will not be a concern.  This condition is satisfied
for the lower redshift simulations.  

For the high redshift simulations, on the other hand, the use of the
pre-COSTAR FOC images presents a concern, as the PSF is quite extended
(the radii for encircled energy fractions of 50\% and 90\% are 38 and 84
pixels, respectively; Leitherer et al.\ 1996).  To account for this, the
UV data were deconvolved in order to recover as much flux as possible
from point sources. 

The UV morphology is dominated by several dozen bright star-forming
knots embedded within a more diffuse background (\cf\ Conti \& Vacca
1994; Meurer et al.\ 1995), and so we chose to deconvolve the images in
two steps.  The first step is to remove as much flux from the point
sources as possible by using the sigma-clean algorithm (Keel 1991).  We
used a large number of iterations (\about 200-400) in order to recover
any significant point sources.  The residual image was then deconvolved
with 10 iterations of the Richardson-Lucy (R-L) deconvolution algorithm. 
This two-step process avoids producing the large ``absorption troughs''
around point sources that result from many iterations of the R-L
algorithm. 

There is also some concern that the limited field of view and
sensitivity of the FOC images prohibit a proper representation of the
far UV morphology of these systems.  This is especially a concern for
the \z=1.59 simulation, where the limiting (2.5$\sigma$) isophote of the
FOC images are detected with a S/N$>$5 in the simulated F606 image. 
However, based on $B-R$ color maps and narrow-band H$\alpha$ images of
these systems (Hibbard, unpublished), we do not expect significant
populations of young UV emitting stars outside of the starburst regions. 
We therefore believe that the UV morphologies are fairly well
represented by the FOC images. 

\section{Construction of Simulated Images}

The $UV, B$, and $V$ images were used to simulate the appearance of the
systems when placed at distances corresponding to redshifts between
\z=0.45 to 2.44 and observed with HST using exposure times similar to
those used for the Hubble Deep Field (HDF; Williams et al.\ 1996).  The
redshifts of the simulated observations are chosen to shift the $UV, B$
and \V-band images into the HST F606W ($\lambda_{pivot}$=5935 \AA) and
F814W ($\lambda_{pivot}$=8269 \AA) filter passbands\footnote{We
calculate the redshifts from the pivot wavelengths of the filter-detector
system, since these are independent of the input spectral energy
distribution.}.  In particular, the \V-image shifts into the HST F814W
passband at \z=0.45, the \B-image shifts into the HST F814W passband at
z=0.81, and the $UV$ image shifts into the HST F606W passband at \z=1.59
and into the F814W passband at \z=2.44. 

A threshold of 2.5$\sigma$ was first applied to the images, after which
they were edited to remove stars, foreground or background galaxies, and
any stray artifacts.  The images were then degraded in surface
brightness and linear scale according to the simulated redshift and the
adopted cosmology.  The technique used to produce the simulations is
similar to that described by Giavalisco et al.\ (1996a), and the
appropriate equations are given in that reference.  We differ from these
authors in the final steps; instead of adding artificial noise to the
images we insert the degraded images directly into a subsection of the
HDF F814W mosaic.  The advantage of our technique is that it ensures
that the detection limit of the simulated images is close to that of the
actual HDF observations and it allows a direct comparison to be made
with HDF images of galaxies at these redshifts. 

We have not considered any evolutionary effects in these simulations;
{\it i.e}, the ages of the underlying populations within the galaxies
are taken to be exactly as they are today.  We do this because we are
interested in assessing the appearance of systems with a very young,
active starburst superposed upon a more evolved underlying population
when viewed from cosmological distances.  In \S5.3 we incorporate
an approximate treatment of evolution and examine its effects.  

The results of the redshifting exercise are shown in Figure 2 [plate YY]
for the intermediate redshift simulations ($z=0.45$ and 0.81), and in
Figure 3 [plate ZZ] for the highest redshift simulations (\z=1.59 and
2.44).  The upper panels present the simulated images of the five
starburst systems (recall that Arp 299 is composed of two separate
systems), while the lower panels show galaxies with similar redshifts
taken from the HDF, displayed at the same spatial scale and with the
same transfer function as the simulated images nearest their redshift. 
The redshifts of the HDF galaxies are listed at the bottom of each
panel, and were taken from the {\it Hawaii Active Catalog of the HDF}
(Cowie 1997) or from Lowenthal et al.\ (1997). 

For redshifts lower than \about 1 the band-shifting is not large enough
to shift rest frame wavelengths shortward of the 4000 \AA\ break into
the simulated passband.  Additionally, the surface brightness
sensitivity of the simulated images is not much worse than in the
ground-based observations.  As a result, the defining morphological
peculiarities of these systems remains readily discernible. 

At the higher redshifts, the rest $UV$ is shifted into the simulated
passband, and the more evolved population is no longer
apparent.\footnote{The \z=1.59 V606 simulation would have a
corresponding F814W observation that measures the rest $U$-band light of
the systems, and it is possible that these observations would detect the
more extended stellar population.  We are testing this by obtaining
$U$-band images of a sample of starburst galaxies, including those used
in this study.} The resulting morphologies are due to the distribution
of bright star forming regions within the galaxies.  These morphologies
range from a single knot (He 2-10), to a knot with surrounding
nebulosity (NGC 1741), to a pair of objects (NGC 1614), and finally to
several closely spaced luminous knots (Arp 299).  This is similar to the
morphologies of high-\z\ galaxies found in the HDF and other deep HST
fields (e.g.~Cowie et al.\ 1995; Abraham et al.\ 1996; van den Bergh et
al.\ 1996; Pascarelle et al.\ 1996; Giavalisco et al.\ 1996b; Koo et
al.\ 1996; Lowenthal 1997). 

We treated our simulated images as if they represented actual
observations of high redshift objects and measured observational
properties (apparent magnitudes, half-light radii, surface brightnesses)
using the same techniques as those used for objects in the HDF
(e.g.~Lowenthal et al.\ 1997).  The results are listed in Table 1.  The
``observed" parameters of the simulated images are quite typical of
those measured for intermediate to high-redshift galaxies (e.g.~Cowie et
al.\ 1995; Koo et al.\ 1996; Pascarelle et al.\ 1996; Giavalisco et al.\
1996a; Lowenthal et al.\ 1997; Trager et al.\ 1997). 
	
In order to estimate the absolute blue magnitude $M_B$, Luminosities,
and SFRs of our simulated objects, we require K-corrections and
conversions between AB magnitudes and magnitudes based on the Vega
system, as well as conversions between the Cousins I band filter and the
HST F814W filter and between the Johnson V filter and the HST F606W
filter.  The conversion between AB magnitudes and Vega-based magnitudes
is simply the difference in the zero points of the magnitude systems and
was taken to be 0.116 mag and 0.439 mag in the F606W and F814W filters,
respectively. 

K-corrections are given by a number of authors for a variety of
redshifts and galaxy types (e.g., Frei \& Gunn 1994; Fukugita, Shimasaku
\& Ichikawa 1995; Kinney et al.\ 1996).  However, K-corrections in the B
band are not available for redshifts as high as $z=2.44$, due primarily
to the fact that observed galaxy spectra (on which the K-corrections are
based) are unavailable at rest wavelengths of $\sim 1200$ \AA.  To
circumvent this problem, we calculated K-corrections for I band
magnitudes by performing synthetic photometry on the low-extinction
spectral energy distribution (SED) template for a starburst (SB1;
E(B-V)$<$0.1) constructed by Kinney et al.\ (1996).  The I band
K-corrections are presented as a function of redshift for the six
starburst templates in Figure 4.  The same starburst templates were used
to calculate the rest frame colors and the conversions between the HST
filter system and the Johnson/Cousins filter system. 

The use of a low-extinction starburst template may seem inappropriate,
given that these systems are known to be dusty.  However, we wish to
treat our simulations as much as possible like actual observations of
high redshift systems, for which a low extinction is usually assumed
(e.g.~Cowie et al.\ 1995; Steidel et al.\ 1996, Lowenthal et al.\ 1997). 
We will examine the effect of this assumption in \S5.2 below. 
 
\section{Discussion}

\subsection{Observed Morphologies}

The redshifting experiment we have performed using nearby starburst
galaxies yields images which look remarkably similar to those of the
faint objects found in the HDF and other deep HST fields.  None of the
previous redshifting experiments reproduce this range of morphologies so
well.  We note that one of the objects in our sample, NGC 1741, is often
used as a {\it spectroscopic} template for galaxies at high redshift
(e.g., Conti et al.\ 1996; Steidel et al.\ 1996).  The similarities in
morphology, SFRs, and spectral energy distributions (\cf~Conti et
al.\ 1996) lead us to conclude that peculiar starburst galaxies appear to
be the best local analogs to the high-$z$ systems. It remains to be
seen if similar experiments using normal types can reproduce {\it all}
of these aspects of the high redshift systems. 

We find that the outer tidal morphologies (and therefore the intrinsic
dynamical nature) of these systems are readily discernible to redshifts
\z\about 1.  This suggests that meaningful comparisons should be
possible between low-\z\ peculiars and those at intermediate redshifts
(\cf~Wu, Faber \& Lauer 1996).  We are in the process of obtaining
ground based \U-band images of a sample of \about 50 peculiar galaxies
to investigate whether this conclusion holds to redshifts of \about 1.3
(Hibbard et al., in preparation).  A comparison between the tidal
morphologies of intermediate redshift peculiars with those of a properly
degraded low-\z\ sample could address the question of whether
interactions have a different nature at the different epochs. 

Figures 2 and 3 illustrate how severe both band-shifting and
cosmological fading can be for $z>$1.5.  At these redshifts most of the
morphologically dominant features in the optical have faded from view,
and only the youngest and most active star forming regions are visible. 
What is quite remarkable is that the dominant morphological features at
these redshifts are {\it not} the most active star forming regions (as
deduced from observations in the far infrared) nor the centers of the
galaxies.  Rather, they are the least obscured star forming regions
visible in the $UV$ images.  For example, in NGC 1614, the main nucleus
(labeled {\bf A} in Fig.~1a) is the site of most of the star formation
activity, while the second nucleus (labeled {\bf B}) has post-burst
spectral characteristics (Liu \& Kennicutt 1995) and a significant
population of UV emitting B-stars.  However, due to very high dust
extinction at the site of region {\bf A}, the optically faint second
nucleus {\bf B} dominates the morphology at high redshifts, with the
main nucleus appearing only as a faint extension to the east. 
Similarly, in Arp 299 the dominant feature at high redshift is not a
galactic nucleus, but the region of disk overlap (region {\bf C} in
Fig.~1b), contributing more than half of the $UV$ photons.  Yet IR and
NIR mapping (Joy et al.\ 1989; Gehrz et al.\ 1982; Wynn-Williams et al.\
1991) suggest that this region contributes only 10\% of the total star
formation activity. 

While it is true that peculiar galaxies at low redshift will appear
peculiar at high redshift, it is important to realize that the reasons
for the apparently peculiar morphologies differ at low and high
redshifts.  At low redshift, the peculiar morphology results from the
action of strong tidal forces on the extended (evolved) stellar
population (\cf~Toomre \& Toomre 1972), and to a lesser extent from the
large number of young stars and large amounts of dust associated with
the merger induced starburst.  At high redshift, the morphology is
driven by the irregular distribution of the most active and/or least
extinguished regions of star formation.

Most of the systems used in this study have an irregular UV morphology
for the same reason that they are classified as peculiar galaxies: they
are on-going mergers of pre-existing systems.  On the other hand, it is
possible that the disturbed UV morphologies at high redshift result
instead from the coalescence of star-forming proto-galactic clumps which
represent the first generation of stars formed in these systems. 
Alternatively, more quiescent scenarios might also give rise to such
morphologies.  For example, galaxies may be assembled in isolation but
with star formation occurring primarily in intense short-lived bursts
that are distributed irregularly throughout an otherwise smooth
distribution of stars and gas (e.g.,~the ``Christmas tree" model of
Lowenthal et al.~1997).  The UV morphology alone cannot distinguish
between these scenarios.  However, the morphology of any extended light
from evolved stars will constrain the true dynamical nature of the
objects, and should discriminate between the various scenarios.

The existence and morphology of an extended evolved population can be
investigated via deep rest-frame optical observations.  While the UV
morphologies at $z$\simgt 1.5 are only \about 1\as\ in diameter, the
stellar disks or spheroidals should be several times larger (5--10 kpc
{\it vs.}~1--2 kpc, \cf~Table 1).  At these redshifts, the optical bands
are shifted into the near-infrared.  Deep, high resolution imaging
observations (e.g., with NICMOS on HST or ground-based NIR observations
under good seeing conditions) should be able to detect irregular
extended distributions such as those in NGC 1741, NGC 1614, and Arp 299,
although they would fail to detect the more regular distribution of
evolved stars from a system like He 2-10.  Such observations should
reveal whether the star forming regions are embedded within a more
diffuse stellar body, or are truly naked starbursts, and hence
proto-galaxies. 

\subsection{Recovery of Global Properties}

In Figure 5 we plot the derived half-light radii, luminosities, and star
formation rates listed in Table 1 as a function of redshift, after
normalizing by their values at $z\approx0$.  This figure shows that the
inferred properties of the high redshift simulations are often quite
different from their actual values.  This illustrates the potential
danger of using restframe UV properties to derive global system
characteristics. 

The reasons for the behavior seen in Fig.~5 are two-fold.  The first is
the presence of an underlying population which is older than the
starburst and contributes significant flux in the optical but very
little in the UV.  The second is the fact that we have made no
correction for dust and have used a low-extinction ($E(B-V)\approx 0.1$)
SED to derive the K-corrections.  While NGC 1741 has a relatively small
intrinsic reddening ($E(B-V)$=0.05; Conti et al.\ 1996), He 2-10 and NGC
3690 are moderately reddened ($E(B-V) \approx$ 0.2--0.3; Vacca \& Conti
1992; Vacca et al.\ 1997), and NGC 1614 and IC 694 are heavily reddened
($E(B-V) \approx$ 1).  In this subsection we examine how these effects
affect the derivation of global values for these systems. 

The behavior of $r_{1/2}(z)$ (Fig.~5a) can be completely understood in
terms of the first of these effects: as the observed band moves into the
rest UV, its morphology is dominated by the most active star forming
regions.  These regions are concentrated within very tight knots or
associations that are considerably more compact than the surrounding
evolved stellar populations.  In all of these systems except He 2-10 the
evolved populations contribute significant optical flux within the
starburst region, and the half-light radius is larger than that of the
young stars alone.  He 2-10 is unresolved in all of our simulations, and
thus the measured values of $r_{1/2}(z)$ are upper limits.  The
resulting behavior of $r_{1/2}(z)$ seen in Fig.~5a is simply due to the
increase in the equivalent physical size of the PSF at higher redshifts. 

The behavior of $L_B(z)$ (Fig.~5b) is a combination of both effects.  In
all cases, the observations at $z<1$ sample the rest optical, and $L_B$
is fairly accurately recovered.  At $z>1$, we must derive \B-band
luminosities from rest UV observations, and both effects are
contributing to the errors.  In addition to taking into account redshift
dilution, a correction of the form $L_B=L_{UV}\times [L_B/L_{UV}]_{SED}$
is required, where $[L_B/L_{UV}]_{SED}$ is calculated from an assumed
template.  Since the UV flux contains no contribution from the more
evolved population, a \B-band luminosity derived from the UV fluxes will
be underestimated.  For the present systems, this amounts to a factor of
two to three.  In addition, if the actual SED is different from that of
the template, the factor $[L_B/L_{UV}]_{SED}$ will be different
(\cf~Fig.~4).  If the underlying galaxy is redder than the template,
$L_B$ will be underestimated, while if it is bluer $L_B$ will be
overestimated. 

For the five systems examined here, NGC 1741 alone is bluer than the
assumed SB1 template.  As a result, the higher $[L_B/L_{UV}]_{SED}$
value partially compensates for the lack of photons from the extended
population, and the derived $L_B$ is coincidentally within 90\% of the
actual value.  For the other four systems, both the presence of an
evolved population and an incorrect ($UV-B$) color cause $L_B$ to be
underestimated, with the luminosities of the reddest systems (NGC 1614,
IC 694) underestimated by the largest factor.  Knowledge of the
reddening in the UV can be incorporated to help alleviate these
shortcomings.  However, such knowledge can never reproduce the
contribution from an underlying evolved population.  For this, the only
recourse is to attempt to observe this population directly in the
restframe optical bands. 

The behavior of $SFR(z)$ seen in Fig.~5c also has contributions from
both effects: the SFR equations have been derived for an unreddened
population with an age of 10 Myrs.  In light of the presence of
significant amounts of dust and substantially older populations in these
systems, these assumptions are not valid.  Therefore the conversions
between broad band luminosities and the SFR given in Table 1 fail to give
consistent results between the various bands. 

Again, knowledge of the reddening in the UV can compensate for some of
these effects.  For example, reddening may be derived by spectroscopy or
by a UV color (e.g.~F606W and F814W observations of $z>2$ galaxies, see
Meurer et al.\ 1997).  However, it should be noted that (1) the colors
so derived will be dominated by relatively unextinguished sightlines,
which may not be characteristic of the system as a whole; and (2) the
correlation between the UV luminosity and the IR luminosity, used to
estimate the total SFR, may not hold for very dusty systems (e.g.,
Meurer et al.\ 1997).  Indeed, NGC 3690 and NGC 1614 lie well off of the
relationship between UV and IR flux given in Meurer et al.\ (1995,
1997).  As a result, the SFRs derived from the IR at $z\approx 0$ are
well above the values derived from the other bands at higher redshift
(Fig.~5c). 

There is a prevailing sentiment that star-forming galaxies at high
redshift are naked starbursts, with very little dust and very few
evolved stars.  Part of this bias is driven by the fact that the
universe is a fraction of its present age at these redshifts (e.g.~2.0
Gyr for \z=2.44 using the adopted cosmology).  However, a number of
systems have been found at $z>1$ that exhibit spectroscopic evidence of
both evolved stellar populations (Stockton, Kellogg \& Ridgeway 1995;
Dunlop et al.\ 1996; Spinrad et al.\ 1997) and copious amounts of dust
(Clements et al.\ 1992; Dunlop et al.~1994).  In addition, the restframe
UV colors of $z>2$ galaxies suggest that reddening in these objects is
significantly higher than normally assumed, with internal color excesses
similar to those found in local starbursts (Meurer et al.\ 1997).  So
the presence of evolved stars and dust should not be discounted {\it a
priori}.  As our observations and simulations have shown, even extremely
dusty galaxies such as NGC 1614 and IC 694 are detectable in the
restframe UV, despite large amounts of reddening.  Presumably, in such
systems the dust is sufficiently patchy in nature that some relatively
unextinguished sightlines exist through which active star-forming
regions can be seen. 

We conclude that if the objects found at high redshift are at all
similar to the galaxies in our sample (i.e.~possessing significant
amounts of dust and/or an evolved population), the sizes, blue
luminosities and SFRs may be substantially underestimated. 

\subsection{Evolutionary Effects}

The simulations we have presented do not include any attempts to model
evolutionary effects in the underlying stellar populations.  That is,
the age of the underlying populations and the level of star formation
throughout the simulated high-\z\ images are presumed to be just as they
are in the low-\z\ analogs.  Our intent is {\em not} to model what these
specific systems looked like when they (and the universe) were younger.
Rather, we simply wish to explore what starbursting peculiar
galaxies containing a range of stellar populations would appear like at
cosmologically interesting distances.  The conclusions arrived at above
do not require the underlying stellar populations be 10 Gyr old, just
that this population be significantly older than the starburst (e.g.~a
few $\times10^8$ years) and contribute significantly to the optical 
luminosity.

It is common to assume that stars form in a galaxy where they are seen
today.  In this case, evolution may crudely be modeled by scaling the
observed \B-band images by a factor to account for the expected increase
in the surface brightness due to a higher star formation rate per unit
area at earlier times, and by the ($UV-B$) color expected for younger
population.  We performed such a simulation assuming an exponentially
decreasing star formation rate with a characteristic time of 2 Gyr
(typical of an Sb galaxy, Bruzual \& Charlot 1993), and formation epoch
of \z=10.  These parameters are used in conjunction with Bruzual \&
Charlot (1997) models to convert the $B$-band images into UV images of
1.7 Gyr old disks, which are then redshifted as before.  The resulting
\z=2.44 simulated images are shown in Figure 6, with the same $z>2$ HDF
systems as shown in the bottom row of Fig.~3. 

As seen in this figure, the HDF galaxies at $z>2$ clearly lack disks as
prominent as those seen in the simulations with our crude treatment of
evolution.  From this, one might conclude that the such disks formed
since z$\sim$2 (e.g., Pascarelle et al.\ 1996).  Alternatively, it is
possible that the assumption of spatially uniform exponential star
formation provides a poor description of disk formation.  Deep NIR
imaging observations should discriminate between these alternatives. 

\section{Conclusions}

We have conducted a redshifting experiment to simulate the appearance 
of local starburst galaxies if observed at cosmological distances.  We
find:

\begin{itemize}

\item{Extended tidal features remain readily visible in the long
exposures typical of the HDF out to redshifts of \about 1.}

\item{For systems above $z\approx$1.5, the simulated morphologies look
remarkably similar to those of the faint objects found in the HDF and
other deep HST fields, more so than previous simulations.}

\item{At $z>1.5$ the observed morphologies reflect the distribution of the
most active and/or least extinguished regions of active star formation. 
The extended distribution of starlight from more evolved stellar
populations is no longer detected.}

\item{Photometric measurements of the $z>1.5$ simulated images fail to
recover the true global properties of the underlying systems.  This is
due to two factors: (1) the inability to image the evolved population
which emits very little in the UV; (2) the assumption that high-\z\
systems are relatively dust free.}

\end{itemize}

If high-\z\ galaxies contain an evolved population of stars which
contribute significantly in the optical but not in UV and/or substantial
amounts of dust, the last two points have particularly important
consequences for our understanding of these systems.  Their combined
effect could lead to erroneous conclusions concerning the nature of
these systems, such as whether they are $L_B^*$ or sub-$L_B^*$ systems,
compact or extended, or have exponential or $r^{1/4}$ light profiles. 
We note that at least some high-redshift objects are known to contain
dust and/or an evolved populations. 

These considerations raise the question of whether the systems recently
discovered at high redshift are indeed ``proto-galaxies",
``proto-spheroids", or ``sub-galactic building blocks".  The galaxies in
our sample are forming a significant population of new stars, and will
likely evolve into dynamically ``hot" stellar systems.  However, they
contain an underlying population of evolved stars and would not be
considered primeval.  We thus conclude that UV morphology alone cannot
reveal whether high-\z\ systems (\z\simgt1.5) are proto-galaxies,
proto-bulges, or starbursts within a pre-existing population. 
Definitive statements regarding the global properties and the true
dynamical nature of these objects require deep imaging observations in
the near-infrared to search for more evolved populations.  Similarly,
whether or not the high redshift systems have dust can be investigated
by deep observations in the restframe IR.  At $z>$2 the far IR bands are
shifted into the sub-mm, and observations using sub-mm bolometers, such
as SCUBA on the JCMT or the planned millimeter array, should provide
crucial confirmation or constraints on the dust content of such systems. 

In presenting these experiments we are not suggesting that high redshift
galaxies are on-going mergers of pre-existing disk galaxies.  However,
as we have demonstrated here, one would be hard pressed to say that this
were {\it not} the case purely on the basis of the available data.  For
now, nearby peculiar starburst galaxies appear to be the best analogs to
the highest redshift galaxies in terms of morphology, star formation
rates and spectral energy distributions.  How appropriate this analogy
is awaits further observations. 

\acknowledgments

We thank Adam Stanford for permission to use the Arp 299 FOC image prior
to publication, and Rogier Windhorst and Jacqueline van Gorkom for
comments on an earlier version of this paper.  JEH acknowledges support
by Grant HF--1059.01--94A from the Space Telescope Science Institute,
which is operated by the Association of Universities for Research in
Astronomy, Inc., under NASA contract NAS5--26555.  WDV acknowledges
support in the form of a Fellowship from the Beatrice Watson Parrent
Foundation.

\newpage
\onecolumn

\figcaption[hibbard.fig1.ps]{Ground-based \B-band images of the five
low-redshift peculiar starburst galaxies used for the redshifting
experiment, displayed with a wrapping greyscale transfer function.  Arp
299 is counted as two separate systems (NGC 3690 and IC 694) in the text
and Table 1.  The region imaged in the archival FOC observations is
outlined.  The 22\as\ field of view of the FOC corresponds to a linear
size of 10.3, 6.6, 8.3 and 1.9 kpc for NGC 1614, Arp 299, NGC 1741 and
He 2-10, respectively (using the redshifts given in Table 1 and H$\_o$=50
km s$^{-1}$ Mpc$^{-1}$).  Several prominent star forming regions in NGC
1614 and Arp 299 are labeled, as are the separate components of HCG 31, of
which NGC 1741 is a member.  \label{hv97.fig1}}

\figcaption[hibbard.fig2.ps]{Simulated observations of the low-\z\
starburst galaxies embedded into the HDF F814W image.  The ground-based
$B$ and $V$ images have been used to simulate HDF observations in the
F814W filter at redshifts of \z=0.45 and 0.81 ($q\_o$=0.5).  The
simulated images were added to random regions of the drizzled HDF image,
and as such occasionally appear close to HDF galaxies.  For example, the
spiral galaxy to the left of the He 2-10 $z$=0.81 simulation is an HDF
galaxy at a $z$=1.15.  The lower panels show eight galaxies from the HDF
at similar redshifts.  Using the convention of naming HDF galaxies by
the WF chip they appear in and the x and y coordinate in version 2 of
the drizzled HDF, these galaxies are (from left to right) top row:
hd4\_1068\_1613, hd4\_0883\_1350, hd3\_0310\_0849, hd2\_0270\_0236; bottom row:
hd3\_1225\_1091, hd2\_1710\_1306, hd4\_1307\_0514, hd4\_1898\_1157.  All
galaxies are shown at the same angular scale, with each box enclosing
8\as\ on a side (54 kpc at \z=0.45 and 66 kpc at \z=0.81), and with the
same transfer function.  This figure shows that the defining
morphological characteristics of the low-\z\ peculiars remain readily
observable out to these redshifts.\label{hv97.fig2}}

\figcaption[hibbard.fig3.ps]{Simulated observations of the low-\z\
starburst galaxies embedded into the HDF F606W and F814W images.  The
archival HST FOC observations have been used to simulate HDF
observations in the F606W filter at a redshift of $z=$1.59 and in the
F814W filter at $z=$2.44.  The lower panels show eight galaxies from the
HDF at similar redshifts and displayed with the same transfer function
as the simulation images.  The HDF galaxies are (from left to right) top
row: hd2\_0993\_1451, hd3\_0306\_1462, hd3\_0988\_1824, hd3\_1506\_0887;
bottom row: hd2\_0649\_1744, hd2\_0666\_1771, hd2\_1160\_1875,
hd2\_0360\_1302.  The panels measure 4\as\ (34 kpc at \z=1.59 and 31 kpc
at \z=2.44) on a side.  At these redshifts the rest UV has been shifted
into the observed passband and none of the extended features from the
evolved stellar population are visible.  \label{hv97.fig3}}

\figcaption[hibbard.fig4.ps]{Cousin's I-band K-corrections as a function
of redshift, determined from the various starburst (SB) templates of
Kinney et al.  (1996).  The K-corrections were combined with rest frame
$(B - I)$ or $(UV - B)$ colors for each template to determine $L\_B$ and
$L\_{UV}$, respectively.  The SB1 template, corresponding to $E(B-V) <
0.1$, was used for all calculations in this paper.  \label{fig4}}

\figcaption[hibbard.fig5.ps]{The logarithms of the {\bf (a)} half-light
radii, {\bf (b)} blue luminosities, and {\bf (c)} star formation rates
derived from the simulated images at different redshifts, all relative
to the values measured in the original images.  At $z>1.5$ the
redshifted images fail to recover the intrinsic measurements from the
host systems.  We conclude that measurements derived from observations
in the restframe UV should be used with caution when attempting to infer
the global properties of the host systems.  \label{fig5}}

\figcaption[hibbard.fig6.ps]{Simulated observations of the low-\z\
starburst galaxies at $z=2.44$, including an approximate treatment of
evolution.  We have assumed an exponential star formation history with a
characteristic time of 2 Gyr, an age of 1.7 Gyr, and a formation epoch
of $z=10$ (see text).  The lower panels show the same four $z>2$ HDF
galaxies as are shown in the lowest panel of Figure 3, now using the
angular resolution and transfer function as the simulation images shown
directly above.  The panels measure 7.5\as\ (58 kpc) on a side.  
\label{hv97.fig6}}
\newpage
\newpage
\input{hv97tab1.tex}

\end{document}

%% file: hv97tab1.tex
\begin{deluxetable}{ccccccccc}
\tablecolumns{9}
\tablewidth{0pc}
\tablecaption{Properties of Redshift Simulations}
\tablehead{
\colhead{(1)}  & \colhead{(2)} & \colhead{(3)} &
\colhead{~~(4)}  & \colhead{(5)} & \colhead{(6)} &
\colhead{(7)}  & \colhead{(8)} & \colhead{(9)} \\
\colhead{$z$}           & \colhead{Filter} & \colhead{$m_z^{AB}$} & 
\colhead{$~~~r_{1/2}$}	& \colhead{$\mu_{r_{1/2}}^{AB}$} & 
\colhead{$M_B$} 	& \colhead{$r_{1/2}$} & 
\colhead{$L_B/L_B^*$}	& \colhead{SFR} \\
\colhead{~}          	& \colhead{~} &
\colhead{(mag)} 	& \colhead{~~~($^{\prime\prime}$)}  & 
\colhead{(mag arcsec$^{-2}$)} & \colhead{(mag)} & 
\colhead{(kpc)} & \colhead{}   & \colhead{($M_{\odot}$ yr$^{-1}$)}}
\startdata
\multicolumn{9}{c}{He 2-10}\\
\hline
0.00291      &   $B$  &   10.7   &   ~~~7.0   &   17.0    & --20.4 &    0.6   & 0.60 & 29 \nl
0.45~~~~     &  I814  &   21.1   &   ~~~0.13  &   18.8    & --20.5 &    1.0   & 0.60 & 29 \nl
0.81~~~~     &  I814  &   22.5   & $<$0.12   &   19.9    & --20.5 &    1.1   & 0.64 & 31 \nl
1.59~~~~     &  V606  &   25.4   & $<$0.12   &   22.7    & --19.8 &    1.1   & 0.32 & 5.0 \nl
2.44~~~~     &  I814  &   26.2   & $<$0.12   &   23.5    & --19.7 &    1.2   & 0.31 & 4.7 \nl
\cutinhead{NGC 1741}
0.0130      &   $B$  &   13.6   &   ~~11.0    &   20.8    & --20.8 &    4.2   & 0.81 & 39 \nl
0.45~~~     &  I814  &   20.8   &   ~~0.61    &   21.7    & --20.8 &    4.6   & 0.82 & 39 \nl
0.81~~~     &  I814  &   22.1   &   ~~0.50    &   22.6    & --20.9 &    4.9   & 0.89 & 43 \nl
1.59~~~     &  V606  &   24.4   &   ~~0.14    &   22.1    & --20.7 &    1.7   & 0.79 & 12 \nl
2.44~~~     &  I814  &   25.2   &   ~~0.16    &   23.2    & --20.7 &    1.9   & 0.75 & 11 \nl
\cutinhead{NGC 3690}
0.0102      &   $B$  &   13.0   &   ~~~8.0    &   19.5    & --20.9 &    2.4   & 0.90 & 120 \nl
0.45~~~     &  I814  &   20.7   &   ~~0.43    &   20.9    & --20.9 &    3.3   & 0.90 & 43 \nl
0.81~~~     &  I814  &   22.1   &   ~~0.27    &   21.2    & --20.9 &    2.7   & 0.92 & 44 \nl
1.59~~~     &  V606  &   25.1   &   ~~0.14    &   22.8    & --20.0 &    1.7   & 0.42 & 6.4 \nl
2.44~~~     &  I814  &   26.0   &   ~~0.14    &   23.7    & --19.9 &    1.8   & 0.37 & 5.6 \nl
\cutinhead{IC 694}
0.0102      &   $B$  &   12.9   &   ~~17.0    &   21.0    & --21.0 &    5.1   & 1.0 & 180 \nl
0.45~~~     &  I814  &   20.4   &   ~~0.78    &   21.8    & --21.2 &    5.9   & 1.2 & 60 \nl
0.81~~~     &  I814  &   22.0   &   ~~0.62    &   22.9    & --21.0 &    6.1   & 1.0 & 49 \nl
1.59~~~     &  V606  &   26.1   &   ~~0.16    &   24.1    & --19.0 &    1.9   & 0.16 & 2.5 \nl
2.44~~~     &  I814  &   27.0   &   ~~0.15    &   24.9    & --18.9 &    1.9   & 0.14 & 2.2 \nl
\cutinhead{NGC 1614}
0.0160      &   $B$  &   13.2   &   ~~10.0    &   20.2    & --21.6 &    4.7   & 1.8 & 220 \nl
0.45~~~     &  I814  &   19.8   &   ~~0.62    &   20.8    & --21.8 &    4.7   & 2.0 & 97 \nl
0.81~~~     &  I814  &   21.4   &   ~~0.52    &   22.0    & --21.6 &    5.1   & 1.7 & 84 \nl
1.59~~~     &  V606  &   25.4   &   ~~0.18    &   23.6    & --19.8 &    2.1   & 0.32 & 5.0 \nl
2.44~~~     &  I814  &   26.0   &   ~~0.20    &   24.5    & --19.9 &    2.4   & 0.35 & 5.4 \nl
\enddata
\end{deluxetable}
\clearpage
\newcounter{n}
Notes to Table 1: 
\begin{list}{n}{\usecounter{n}\setlength{\rightmargin}{\leftmargin}}

\item[(1)] Redshift 
\item[(2)] Filter for measured values.
\item[(3)] Apparent magnitude, in AB magnitudes.
\item[(4)] Half light radii from aperture photometry, in arcsec. 
\item[(5)] Surface brightness averaged over the half-light radius, in AB magnitudes.
\item[(6)] Absolute blue magnitude calculated for a cosmology with 
$H_o$=50 km s$^{-1}$ Mpc$^{-1}$ and  $q_o$=1/2 according to the following equation: 
$$M_B = m^{AB}_{i}(z) - 5\times log(D_L/10 {\rm pc}) +
[ m_{i}(z) - m^{AB}_{i}(z) ]_{SED} +
[ m_B(0) - m_{i}(0) ]_{SED} - K_{i}$$
where $m^{AB}_{i}(z)$ is the measured magnitude in the AB system, 
$D_L$ is the luminosity distance in pc,
$K_{i}$ is the K-correction for the observed filter $i$ listed in column 2 (see 
Fig.~4 for K-corrections for the F814W filter), $m_i(0)$ is the apparent 
magnitude in filter $i$ at $z=0$, 
and the SED subscript indicates that 
the values are calculated by integrating the chosen SED (SB1 from 
Kinney et al.~1995 in this case) over the appropriate bandpass. Magnitudes are 
in the Vega-based system unless designated with an AB superscript.
\item[(7)] Half-light radii in kpc, using the adopted cosmology.
\item[(8)] Blue Luminosity in terms of $L_B^*$, where we adopt 
$M_B^*\approx$-21 as the characteristic luminosity
for a Schechter luminosity function from Loveday et al.~1992. 
\item[(9)] Star formation rate, derived from
the Bruzual \& Charlot (1997) models, with the assumptions of a Salpeter
Initial Mass Function spanning the mass range 0.1--125 M$_\odot$,
constant star formation, and a population age of 10 Myrs.  This yields
the following equations for the star formation rate in M$_\odot$ yr$^{-1}$:
\begin{eqnarray}
{\rm SFR} & = & L_{bol} / 4.40 \times 10^9 L_\odot~~,\nonumber \\
{\rm SFR} & = & L_B / 4.11 \times 10^{38} {\rm ergs\, s}^{-1} {\rm \AA}^{-1}~~,\nonumber\\
{\rm SFR} & = & L_{2200} / 1.87 \times 10^{39} {\rm ergs\, s}^{-1} {\rm \AA}^{-1}~~.\nonumber
\end{eqnarray}
At $z\approx 0$ the SFRs for He 2-10 and NGC 1741 are calculated from the
observed $B$-band luminosity, while for NGC 3690, NGC 1614 and IC 694 the SFRs are
calculated assuming $L_{bol}\approx L_{IR}$. $L_{IR}$ is derived from 
a single color fit to the four IRAS passbands (Sanders \& Mirabel 1996). At
$z= 0.45, 0.81$ the SFRs are calculated via the derived $L_B$; at $z=1.59, 
2.44$ they are calculated from ${L_{2200}}$. 
\end{list}